\RequirePackage{lineno}
\documentclass[aps,prc,twocolumn,showpacs,superscriptaddress,groupedaddress,nofootinbib]{revtex4}  
\usepackage{graphicx}  
\usepackage{amssymb}
\usepackage{subfigure}
\usepackage{multirow}

\begin{document}
%\leftline{ \today}

\title{Centrality dependence of subthreshold $\phi$ meson production in Ni+Ni collisions at 1.9A GeV}
% Force line breaks with \\
%\thanks{krzysztof.piasecki@fuw.edu.pl}%

\def\wars{Institute of Experimental Physics, Faculty of Physics, University of Warsaw, Warsaw, Poland}
\def\ncbj{National Centre for Nuclear Research, Otwock, Poland}
\def\heid{Physikalisches Institut der Universit\"{a}t Heidelberg, Heidelberg, Germany}
\def\darm{GSI Helmholtzzentrum f\"{u}r Schwerionenforschung GmbH, Darmstadt, Germany}
\def\seou{Korea University, Seoul, Korea}
\def\cler{Laboratoire de Physique Corpusculaire, IN2P3/CNRS, and Universit\'{e} Blaise Pascal, Clermont-Ferrand, France}
\def\zagr{Ru{d\llap{\raise 1.22ex\hbox{\vrule height 0.09ex width 0.2em}}\rlap{\raise 1.22ex\hbox{\vrule height 0.09ex width 0.06em}}}er Bo\v{s}kovi\'{c} Institute, Zagreb, Croatia}
\def\munI{Excellence Cluster Universe, Technische Universit\"{a}t M\"{u}nchen, Garching, Germany}
\def\munII{E12, Physik Department, Technische Universit\"{a}t M\"{u}nchen, Garching, Germany}
\def\vien{Stefan-Meyer-Institut f\"{u}r subatomare Physik, \"{O}sterreichische Akademie der Wissenschaften, Wien, Austria}
\def\sp{University of Split, Split, Croatia}
\def\buda{Wigner RCP, RMKI, Budapest, Hungary}
\def\mosc{Institute for Theoretical and Experimental Physics, Moscow, Russia}
\def\dres{Institut f\"{u}r Strahlenphysik, Helmholtz-Zentrum Dresden-Rossendorf, Dresden, Germany} 
\def\harb{Harbin Institute of Technology, Harbin, China}
\def\kurc{National Research Centre "Kurchatov Institute", Moscow, Russia}
\def\buch{Institute for Nuclear Physics and Engineering, Bucharest, Romania}
\def\stra{Institut Pluridisciplinaire Hubert Curien and Universit\'{e} de Strasbourg, Strasbourg, France}
\def\tsing{Department of Physics, Tsinghua University, Beijing 100084, China}
\def\lan{Institute of Modern Physics, Chinese Academy of Sciences, Lanzhou, China}

\author{K.~Piasecki}\email{krzysztof.piasecki@fuw.edu.pl} \affiliation{\wars}
\author{Z.~Tymi\'{n}ski} \affiliation{\wars} \affiliation{\ncbj}
\author{N.~Herrmann} \affiliation{\heid}
\author{R.~Averbeck} \affiliation{\darm}
\author{A.~Andronic} \affiliation{\darm} 
\author{V.~Barret} \affiliation{\cler} 
\author{Z.~Basrak} \affiliation{\zagr} 
\author{N.~Bastid} \affiliation{\cler}
\author{M.L.~Benabderrahmane} \affiliation{\heid}
\author{M.~Berger} \affiliation{\munI} \affiliation{\munII}
\author{P.~Buehler} \affiliation{\vien} 
\author{M.~Cargnelli} \affiliation{\vien} 
\author{R.~\v{C}aplar} \affiliation{\zagr}
\author{E.~Cordier} \affiliation{\heid}
\author{P.~Crochet} \affiliation{\cler} 
\author{O.~Czerwiakowa} \affiliation{\wars}
\author{I.~Deppner} \affiliation{\heid}
\author{P.~Dupieux} \affiliation{\cler}
\author{M.~D\v{z}elalija} \affiliation{\sp}
\author{L.~Fabbietti} \affiliation{\munI} \affiliation{\munII}
\author{Z.~Fodor} \affiliation{\buda}
\author{P.~Gasik} \affiliation{\wars} \affiliation{\munI} \affiliation{\munII}
\author{I.~Ga\v{s}pari\'c} \affiliation{\zagr}
\author{Y.~Grishkin} \affiliation{\mosc}
\author{O.N.~Hartmann} \affiliation{\darm}
\author{K.D.~Hildenbrand} \affiliation{\darm}
\author{B.~Hong} \affiliation{\seou}
\author{T.I.~Kang} \affiliation{\seou}
\author{J.~Kecskemeti} \affiliation{\buda}
\author{Y.J.~Kim} \affiliation{\darm}
\author{M.~Kirejczyk} \affiliation{\wars} \affiliation{\ncbj}
\author{M.~Ki\v{s}} \affiliation{\darm} \affiliation{\zagr}
\author{P.~Koczon} \affiliation{\darm}
\author{M.~Korolija} \affiliation{\zagr}
\author{R.~Kotte} \affiliation{\dres}
\author{A.~Lebedev} \affiliation{\mosc}
\author{Y.~Leifels} \affiliation{\darm}
\author{A.~Le F\`{e}vre} \affiliation{\darm}
\author{J.L.~Liu} \affiliation{\heid} \affiliation{\harb}
\author{X.~Lopez} \affiliation{\cler}
\author{A.~Mangiarotti} \affiliation{\heid}
\author{V.~Manko} \affiliation{\kurc}
\author{J.~Marton} \affiliation{\vien}
\author{T.~Matulewicz} \affiliation{\wars}
\author{M.~Merschmeyer} \affiliation{\heid}
\author{R.~M\"{u}nzer} \affiliation{\munI} \affiliation{\munII}
\author{D.~Pelte} \affiliation{\heid}
\author{M.~Petrovici} \affiliation{\buch}
\author{F.~Rami} \affiliation{\stra}
\author{A.~Reischl} \affiliation{\heid}
\author{W.~Reisdorf} \affiliation{\darm}
\author{M.S.~Ryu} \affiliation{\seou}
\author{P.~Schmidt} \affiliation{\vien}
\author{A.~Sch\"{u}ttauf} \affiliation{\darm}
\author{Z.~Seres} \affiliation{\buda}
\author{B.~Sikora} \affiliation{\wars}
\author{K.S.~Sim} \affiliation{\seou}
\author{V.~Simion} \affiliation{\buch}
\author{K.~Siwek-Wilczy\'{n}ska} \affiliation{\wars}
\author{V.~Smolyankin} \affiliation{\mosc}
\author{G.~Stoicea} \affiliation{\buch}
\author{K.~Suzuki} \affiliation{\vien}
\author{P.~Wagner} \affiliation{\stra}
\author{I.~Weber} \affiliation{\sp} 
\author{E.~Widmann} \affiliation{\vien}
\author{K.~Wi\'{s}niewski} \affiliation{\wars} \affiliation{\heid} 
\author{Z.G.~Xiao} \affiliation{\tsing}
\author{H.S.~Xu} \affiliation{\lan}
\author{I.~Yushmanov} \affiliation{\kurc}
\author{Y.~Zhang} \affiliation{\lan}
\author{A.~Zhilin} \affiliation{\mosc}
\author{V.~Zinyuk} \affiliation{\heid}
\author{J.~Zmeskal} \affiliation{\vien}

\collaboration{FOPI Collaboration} \noaffiliation

\date{\today}% It is always \today, today,

\begin{abstract}

We analysed the $\phi$ meson production in central Ni+Ni collisions 
at the beam kinetic energy of 1.93A GeV with the FOPI spectrometer
and found the production probability per event of
$[8.6 ~\pm~ 1.6 ~(\text{stat}) \pm 1.5 ~(\text{syst})] \times 10^{-4}$. 
This new data point 
allows for the first time to inspect the centrality dependence 
of the subthreshold $\phi$ meson production in heavy-ion collisions.
The rise of $\phi$ meson multiplicity per event with mean number 
of participants can be parameterized by the power function 
with exponent $\alpha = 1.8 \pm 0.6$. 
The ratio of $\phi$ to $\text{K}^-$ production yields
seems not to depend within the experimental uncertainties 
on the collision centrality,
and the average of measured values was found to be $0.36 \pm 0.05$. 

\end{abstract}

\maketitle
%\linenumbers

\section{Introduction}
\label{Sect:intro}

The $\phi$ meson is a particularly interesting hadron for a variety of reasons. 
It is a non-strange particle composed of a $\text{s}\bar{\text{s}}$ pair of quarks,
and is characterized by the narrow mass distribution
centered at $m_{\phi} = 1.0195$~GeV/$c^2$, and decay width
$\Gamma = 4.27$~MeV~\cite{PDG}.
Its suppressed decay modes have accompanied the discovery of the 
Okubo-Zweig-Iizuka (OZI) rule~\cite{OZI}.
An enhancement of its production in the ultra-relativistic heavy-ion (AA)
collisions has been interpreted as the result of fragmentation of gluons
into q$\bar{\text{q}}$ pairs in the Quark-Gluon Plasma~\cite{Koch90}. 
Calculations in the frame of the transport BUU code suggest that $\phi$ mesons 
from the heavy-ion collisions at energies below threshold in a free 
nucleon-nucleon (NN) collision may be produced on average even earlier
than kaons, and due to relatively weak absorption and rescattering 
cross sections, a majority of $\phi$'s may survive the collision. 
Therefore, $\phi$ mesons may be a good probe of earlier stages of 
the heavy-ion collision~\cite{Scha10}. 

In the proton-proton (pp) collisions, the ratio of $\phi$ to $\omega$ 
cross sections obtained near $\phi$ meson threshold was found 
to violate the OZI rule by about an order of magnitude~\cite{Bale01,Hart06}. 
Studies of near-threshold $\phi$ meson production in pA collisions 
revealed the target mass dependence of A$^{0.56(3)}$~\cite{Taba06}, 
and pointed to the momentum-dependent widening of the $\phi$ width 
in the nuclear medium, from where the $\phi$N absorption cross 
section of about 15--20~mb was estimated~\cite{Hart12}.

In the domain of heavy-ion collisions data on the subthreshold $\phi$ meson 
production is quite scarce. In the first measurement only $23 \pm 7$
events were found in the K$^+$K$^-$ channel
and the reconstructed yield per event varied by the \mbox{factor 4}
depending on the value of temperature parameter of the $\phi$ 
meson source assumed for the efficiency evaluation~\cite{Mang03}. 
Recently, three samples of about \mbox{110--170} $\phi$ mesons were measured 
at quite similar values of the average number of participant nucleons 
($\langle A_{\text{part}} \rangle_{\text{b}}$ = \mbox{35--50}), 
and beam kinetic energies in the range of 1.75--1.9A GeV
~\cite{Agak09,Lore10,Pias15,Gasi16}. Production yields per triggered event 
were found to be $2.5 \text{--} 4.5 \times 10^{-4}$. 
Taking into account that the branching ratio equals
$\text{BR} (\phi \rightarrow \text{K}^+\text{K}^-$) = 48.9 \%~\cite{PDG},
and comparing to the corresponding production yields of K$^-$, 
it was inferred that $\phi$ mesons were the source of about 
\mbox{15--20~\%} of negatively charged kaons. 
A determination of the size of this contribution,
and tracing its dependency on the beam energy and collision centrality
is important in the context of the investigation of in-medium modifications 
of K$^-$ mesons from their emission patterns~\cite{Fuch06,Lutz04,Hart11,Wisn00,Fors07}. 
As the mean decay path of $\phi$ meson is 46~fm, it decays 
mostly outside the collision zone, and therefore the majority of 
daughter kaons are not produced in the medium. 
Phase space distribution of negative kaons is thus composed
of two contributions, and the extraction procedure of 
the kaon potential inside the medium should take into account
the side feeding from $\phi$ meson decays. 

In recent years there has been a considerable development 
on the side of transport models. 
Earlier calculations in the frame of the BUU code 
performed for the collisions of central Ni+Ni at 1.93A GeV, 
and Ru+Ru at 1.69A GeV
favored the dominance of meson-baryon production channels 
($\text{MB} \rightarrow \phi \text{B}$, M = [$\rho$,~$\pi$], 
B = [N,~$\Delta$])~\cite{Barz02}.
However, further BUU calculations employing the new effective 
parameterization of the
$\text{pN} \rightarrow \text{pN} \phi$ channels (N = [p,n]), 
performed for the central Ni (1.93A GeV) + Ni and semi-central 
Ar (1.756A GeV) + KCl collisions
suggested the BB production channels to be of equal strength as MB.
An interesting study of $\phi$ meson production in Ca (1.76A GeV) + Ca 
collisions in the frame of the UrQMD model has been recently published~\cite{Stei16}. 
Within this approach the decays of massive resonances 
[$N^*(1990) ~...~ N^*(2250)$] have been pointed out as a 
dominant source of $\phi$ mesons (cf. Fig.~6 therein). 
In both types of calculations the \mbox{$\phi\slash\text{K}^-$} 
ratio was predicted to be similar to the experimental findings 
in the Ar+KCl and Ni+Ni systems.
The BUU model gave detailed predictions on the composition of production channels 
as a function of impact parameter, distribution of production times 
and kinematic variables, while the UrQMD calculations predicted a specific 
profile of the excitation function of \mbox{$\phi\slash\text{K}^-$} ratio 
with a maximum around the threshold energy, 
evolving around the AGS beam energy region into the constant, lower value. 
Thermal models also deliver predictions of the $\phi$ meson 
production yields in an event.
In Ref.~\cite{Cley99} the \mbox{$\phi\slash\pi^+$} ratio is predicted 
to be centrality independent,
and correlated only with the temperature of the system.
Calculations of the \mbox{$\phi\slash\text{K}^-$} ratio shown in Fig.~13 of 
Ref.~\cite{Agak09} predict the strong enhancement around the threshold, 
sensitive to the suppression of the volume in which the open strangeness 
is produced ($R_{\text{C}}$ parameter). 

However, there seems to be a strong disparity between the variety 
of predictions of subthreshold $\phi$ meson production 
and the scarcity of available experimental data. 

In our work we address this gap by analysing $\phi$ mesons produced 
in the central Ni+Ni collisions at the beam kinetic energy of 1.93A GeV,
and decaying in the dominant K$^+$K$^-$ channel. 
While, as mentioned above, these mesons were already investigated 
within the similar centrality class, 
a relatively low number of $4.7 \times 10^6$ acquired events 
in the experiment reported in~\cite{Mang03}
resulted in only $23 \pm 7$ events attributed to $\phi$ mesons. 
In the experiment described in this paper, about 17 times more events
were acquired for the central trigger (defined in Sect.~\ref{Sect:exp}), 
giving a good chance to measure $\phi$ mesons at considerably 
better significance level. 

As data on $\phi$ meson production at 1.9A GeV was also published
for the less central Ni+Ni collisions, as well as central Al+Al collisions, 
our investigation allows for the first time to gain insight into 
the centrality dependence of the subthreshold $\phi$ meson production 
in heavy-ion collisions.

\section{Experiment}
\label{Sect:exp}

The S261 experiment was performed at the SIS--18 synchrotron in the GSI, Darmstadt.
$^{58}$Ni ions were accelerated to the kinetic energy of 1.93A GeV,
and were incident on the $^{58}$Ni target of the FOPI modular spectrometer
with the average intensity of $4\text{--}5 \times 10^5$ ions per spill. 
The target had thickness of 360 mg/cm$^2$ 
corresponding to 1.0\% interaction probability.

The innermost detector of the FOPI setup is the
Central Drift Chamber (CDC) covering the wide range 
of polar angles ($27^\circ < \theta_{\text{lab}} < 113^\circ$, 
\footnote{The angles are given with respect to the target position 
in the S261 experiment.}), 
and subdivided azimuthally into 16 identical sectors.
The CDC was surrounded by the Time-of-Flight (ToF) detector named
Plastic Scintillation Barrel (PSB), which
covered the polar angles of $26.5^\circ < \theta_{\text{lab}} < 56^\circ$. 
Both detectors were mounted inside the magnet solenoid, 
which generated the field of 0.6~T. The Plastic scintillation
Wall (PlaWa) was placed downstream of the CDC.
More information on the geometry and performance of the FOPI apparatus 
can be found in~\cite{FOPI}. 

The on-line "central" trigger was based on the large number of hits
in the PlaWa.
After an additional off-line removal of events with vertex position
outside the target, about $79\times 10^6$ events were selected,
amounting to $(22 \pm 1)$\% of the total reaction cross-section. 
Assuming the sharp cut-off approximation between 
the total reaction cross section and maximum impact parameter, 
and the geometrical model of interpenetrating spheres, 
the number of participant nucleons averaged over the impact parameter 
was estimated to be $\langle A_{\text{part}} \rangle_{\text{b}} = 74 \pm 2$. 

\section{Data analysis}
\label{Sect:ana}

Charged particles passing through the CDC detector
activate the nearest sense wires, 
represented within the data acquisition scheme as "hits". 
An off-line tracking routine collects them into "tracks", 
and reconstructs the emission angles and specific energy loss of particles. 
For these particles which additionally generate hits in the PSB detector,
the ToF information is also obtained. 
A reconstruction of the collision vertex allows for the elimination of
events not originating from the target. 

To preselect the tracks with good reconstruction quality, 
these composed of few hits were rejected.
The tracks with too large distance of closest
approach to the vertex were also eliminated. 
The candidates for K$^+$ and K$^-$ mesons were searched 
from the tracks matched with hits in the ToF barrel. 
For a good quality of matching, the extrapolation of the 
CDC track was required to pass close to the hit candidate
in the PSB detector. 
To minimize the edge effects, the range of accepted polar angles 
was trimmed down to $30^\circ < \theta_{\text{lab}} < 53^\circ$,
and K$^+$ (K$^-$) candidates were required to have the transverse
momentum $p_{\text{T}}$ of at least 0.18 (0.14) GeV/$c$. 
A motivation for these cuts will be also discussed in Sect.~\ref{Sect:eff}.

\begin{figure}[tbh]
 \includegraphics[width=8.6cm]{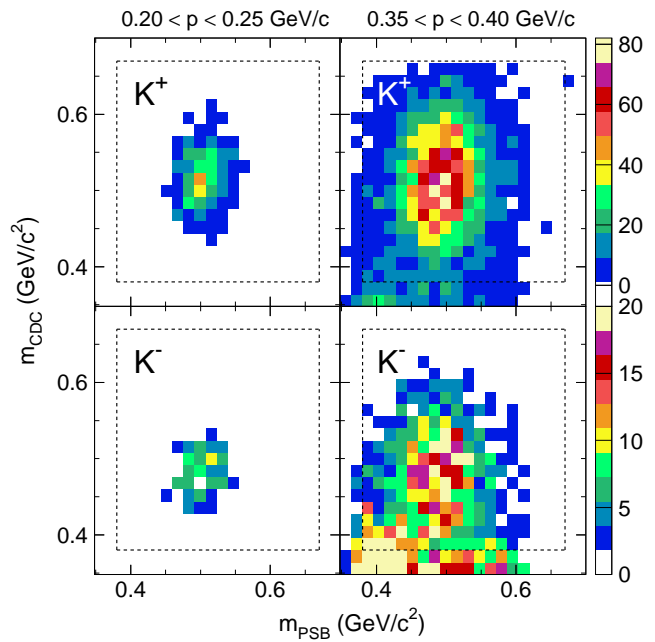}
 \caption{\label{fig:kpkmmass}(Color online)
  Identification of charged kaons on the
  two-dimensional map of FOPI mass parameters: $m_{\text{CDC}}$ 
  (obtained from the CDC) versus $m_{\text{PSB}}$
  (from the combination of CDC-PSB data). Rectangles in dashed lines
  delineate the cuts imposed for the selection of K$^+$ and K$^-$ candidates.}
\end{figure}

Within the basic identification technique of charged particles in 
FOPI the information from a drift chamber (here: the CDC) is used. 
A sign of electric charge is inferred from the direction 
of the track curvature on the transverse plane. 
Subsequently, the momentum and specific energy loss inside the chamber
is substituted to the Bethe-Bloch formula, and the mass parameter
is extracted (dubbed here $m_{\text{CDC}}$). 
However, the resolution of the specific energy loss is too weak for
the identification of kaons in a reasonable momentum range.
Therefore, an additional identification technique is applied:
the mass is reconstructed by substituting the momentum and velocity 
to the relativistic formula $p = m \gamma v$. The mass parameter
obtained in this method is dubbed $m_{\text{PSB}}$. 
For the present analysis both methods were applied together. 
In order to reject most of the pion and baryon events, the gates 
were imposed on the abovementioned mass parameters, 
shown as dashed rectangles in Fig.~\ref{fig:kpkmmass}.

The $\phi$ mesons in the K$^+$K$^-$ decay channel were investigated 
in the invariant mass ($M_{\text{inv}}$) distribution of kaon pairs. 
As the resolutions of $m_{\text{CDC}}$ and $m_{\text{PSB}}$ distributions
in FOPI are well known to deteriorate with momentum,
to limit the influence of particles with neighbouring masses 
(pions and protons), the maximum momentum was set to 0.65~GeV/$c$. 
In the FOPI analyses aiming at the selection of K$^+$ and/or K$^-$
using PSB, a reasonable level of signal to background 
(S/B) ratio was usually kept if particles were limited to lower 
maximum momenta: about 0.55~GeV/$c$ for K$^+$, 
and 0.38~GeV/$c$ for K$^-$~\cite{Wisn00,Ziny14,Pias15,Gasi16}. 
However, for the $\phi$ meson reconstruction analysis an unambiguous
selection of kaons is not so critical, as some admixture of misidentified 
particles is not expected to create a correlation around $m_{\phi}$.
%It has been verified that the statistical significance of the signal,
%and quality of its profile does not deteriorate if the maximum momentum 
%of K$^+$ and K$^-$ is raised to 0.65~GeV/$c$ (this approach has been already
%succesfully applied e.g. in~\cite{Mang03,Pias15,Gasi16}). 
To minimize possible distortions arising for the kaon pair candidates with 
small relative angles, pairs were required not to intersect in the active 
region of the CDC. 
The background of uncorrelated pairs was reconstructed using
the mixed events method, where K$^+$ and K$^-$ tracks from different 
events were matched. The events from where two kaons were paired
had to be attributed to the same centrality class, 
determined by the multiplicity of tracks in the CDC. 
To prevent from smearing out of possible flow pattern of kaons
by matching events with randomly oriented reaction planes,
the parent events were rotated azimuthally to align these planes.
Finally, the background was normalized to the true pair distribution 
in the region $1.05 < M_{\text{inv}} < 1.18$~GeV/$c^2$.

\begin{figure}[tbh]
 \includegraphics[width=8.6cm]{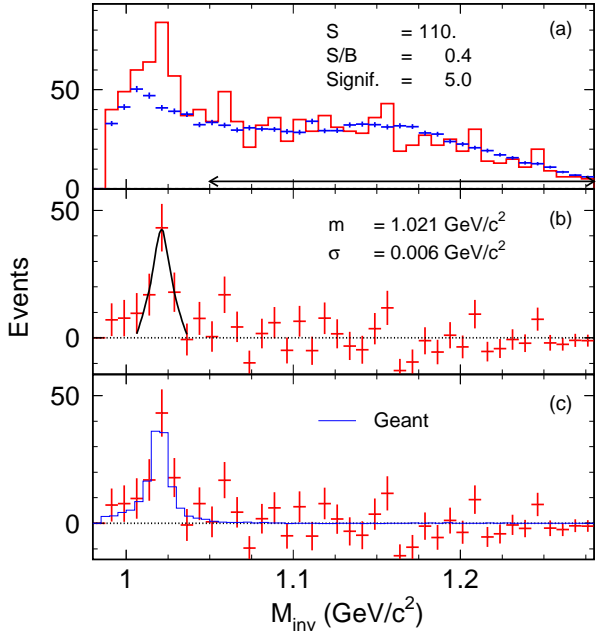}
 \caption{\label{fig:phiminv}(Color online)
   (a) Invariant mass plot of true (solid line), 
   and mixed (scattered points) K$^+$K$^-$ pairs. The normalisation window
   for the spectrum of mixed pairs is indicated with the arrow.
   (b) $\phi$ meson signal obtained after the background subtraction. 
   The solid curve shows the Gaussian distribution fitted within 
   $1.00 < M_{\text{inv}} < 1.04$~GeV/$c^2$.
   (c) A comparison of profiles of experimental (points) and simulated 
   (solid line) $\phi$ meson signals.}
\end{figure}

The distribution of true and mixed pairs is shown in 
Fig.~\ref{fig:phiminv}(a). The $\phi$ meson signal obtained
by the subtraction of background from the spectrum of true pairs is
presented in panel (b) and exhibits some excess of counts in the region
$M_{\text{inv}} < 1.01$~GeV/$c^2$. To check if this excess
is generated by true $\phi$ meson events or if it is a side effect,
the experimental signal was compared to the profile obtained
in the simulation of $\phi$ meson production and detection,
shown by the solid line in Fig.~\ref{fig:phiminv}(c). 
Here we only report the final result of this simulation, 
as it will be described in details further in Sect.~\ref{Sect:eff}. 
As both distributions exhibit the abovementioned excess,
the $\phi$ meson signal was interpreted as the total 
number of signal counts in the range $0.98 < M_{\text{inv}} < 1.05$~GeV/$c^2$.
Within the set of cuts described above, 
the number of identified $\phi$ mesons was found to be $110 \pm 19$.
In addition, a Gaussian distribution was fitted to the signal 
in the range of $1.00 < M_{\text{inv}} < 1.04$~GeV/$c^2$,
yielding the $\sigma$ parameter of 6~MeV/$c^2$
(see Fig.~\ref{fig:phiminv}(b)). 
The peak maximum was found to be $1.0211(13)$~GeV/$c^2$,
about 1 standard deviation away from the nominal value in vacuum. 
Within the range of $\pm ~ 2\sigma$ the signal to background ratio 
was found to be 0.4, and the significance 5.0. 
The number of reconstructed $\phi$ mesons and the other 
characteristics of $M_{\text{inv}}$ distribution depend on the choice 
of cuts applied to the track quality parameters, mass parameters,
maximum momenta, binning, and windows for integration and normalization.
This issue is further discussed in Appendix~\ref{Sect:detaSystErr}.

\section{Efficiency determination}
\label{Sect:eff}

The calculation of efficiency was performed using the GEANT3-based
environment~\cite{GEANT}, where the modeled detectors were
positioned as in the S261 experiment. \mbox{$\phi$ mesons} were sampled
from the relativistic Breit-Wigner mass profile with the nominal mass 
and decay constant~\cite{PDG}, and the phase space was populated 
according to the Boltzmann function multiplied by the 
simple anisotropy term:
\begin{equation}
 \label{EQ:phigeantmodel}
 \frac{d^2N}{dEd\theta} \sim ~pE~ \exp (-E\slash T_{\text{s}})  \cdot \left(1 + a_2 \cos^2 \theta\right) ,
\end{equation}
\noindent where $T_{\text{s}}$ is the temperature of the source, 
and $a_2$ is the anisotropy parameter. Little is known about
the characteristics of the $\phi$ meson distribution in phase space. 
The inverse slopes for the collisions at 1.9A GeV, and lower 
$\langle A_{\text{part}} \rangle_{\text{b}}$, were found to be
between 70 and 130 MeV~\cite{Pias15,Gasi16}. 
As for baryons and K$^\pm$ mesons more central collisions translate
into somewhat higher inverse slopes~\cite{Hong97,Fors07}, 
one may cautiously apply this trend to $\phi$ mesons. 
Thus, for the determination of the efficiency in the present analysis
$T_{\text{s}}$ was varied between 80 and 140 MeV.
Regarding the polar anisotropy of $\phi$ meson emission,
no data is known so far. 
Therefore, as roughly expected from the systematics for other strange 
particles produced in the studied beam energy range and collision
centrality~\cite{Fors07}, $a_2$ was varied between 0 and 1.
The generated $\phi$ mesons were boosted to the laboratory frame,
and all of them were allowed to decay into K$^+$K$^-$ pairs 
(the BR factor was accounted for in the calculation of the efficiency). 
Particles were subsequently added to the events
of Ni+Ni collisions generated by the IQMD transport code~\cite{IQMD}, 
which aims at providing the realistic set of particles emitted 
from the heavy-ion collisions. However, kaons emitted directly from the
collision zone were not produced by this code. 

After the propagation and digitization stages performed by the GEANT routines,
events were processed using the same tracking, matching and 
correlation routines, as for the experimental data. 
Distributions of the quality parameters for the experimental and simulated
tracks were found to mostly overlap. To account for minor discrepancies,
the cut values for the simulated tracks were tuned such that 
a given cut filtered out the same fraction of tracks in the
generated sample, and in the true (measured) one. 
However, the distributions of mass parameters for the 
tracks in the simulation were found to be narrower 
than these from the experiment. 
To tackle this problem, in the first step
for the experimental tracks of K$^+$ and K$^-$ candidates
the two-dimensional distributions of $m_{\text{CDC}}$ versus $m_{\text{PSB}}$
were inspected for consecutive slices of momentum,
as shown in Fig.~\ref{fig:kpkmmass}. 
In the next step, two-dimensional Gaussian functions were fitted 
to these distributions within regions not influenced 
by other particles (pions and/or protons). 
A comparison of the extracted Gaussian distribution
to the rectangular cuts applied on the $m_{\text{CDC}}-m_{\text{PSB}}$ 
plane (see dashed rectangles in the abovementioned Figure)
allowed to find the fraction $f(p)$ of kaons 
rejected by these cuts. Depending on momentum and particle sign,
$f(p)$ ranged from 0.01 to 0.15. 
On the side of the simulation, the narrower mass spectra allowed to 
select the sample of kaons nearly unambiguously. However, to reproduce 
the abovementioned loss of kaons occuring for the experimental data,
a fraction $f(p)$ of simulated kaons was removed from the sample.

The obtained profile of the invariant mass spectrum of K$^+$K$^-$ 
pairs is shown as solid line in Fig.~\ref{fig:phiminv}(c).
The \mbox{$\phi$ meson} reconstruction efficiency in the K$^+$K$^-$ decay channel,
calculated for the set of cuts described above, and for the source parameters
$T_{\text{s}} = 120$~MeV, and $a_2 = 0$ was found to be $3.5 \times 10^{-3}$. 
It was also found that the efficiency obtained with $a_2 = 1.0$ 
was only 2\% lower, while variations due to different assumed values of
$T_{\text{s}}$ remained within $\pm 10 \%$ range.

In the next step we have checked whether the simulation reproduced
the efficiency of matching of tracks in the CDC with hits in the PSB.
As the statistics on kaons was too low for a reliable 
study of this issue, the effect for negatively charged particles
was tested on $\pi^-$ tracks, and that for positively charged ones
-- on the sum of $\pi^+$ and protons. 
In the first step the ratio of CDC tracks with associated 
hit in the PSB to all the reconstructed CDC tracks was inspected 
as a function of laboratory polar angle and transverse momentum. 
This ratio was obtained independently for the experimental 
and simulated data. Subsequently, the ratio obtained for 
the experimental data was divided by that for the simulated data, 
to yield the correction factor for the matching efficiency,

\begin{equation}
\varepsilon^{CDC-PSB} \left( \theta_{\text{lab}}, p_{\text{t}} \right) 
 = \frac{N^{\text{PSB}}_{\text{exp}}}{N^{\text{CDC}}_{\text{exp}}}
   \slash
   \frac{N^{\text{PSB}}_{\text{sim}}}{N^{\text{CDC}}_{\text{sim}}} \quad .
\end{equation}

\begin{figure}[bt]
 \includegraphics[width=8.6cm]{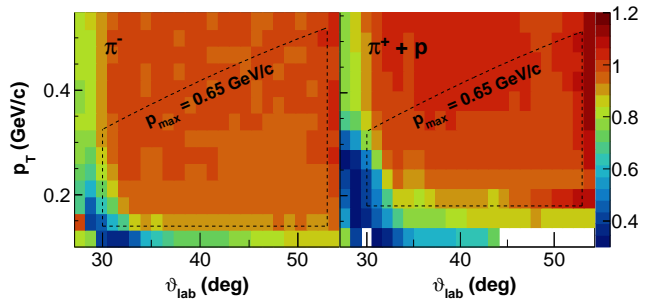}
 \caption{\label{fig:bareff}Maps of correction factor for possible 
unequal efficiencies of matching of CDC tracks with PSB hits between 
the experimental and simulated data,
obtained for (a) $\pi^-$ tracks, and (b) sum of $\pi^+$ and proton tracks.
Regions in between dashed lines correspond to cuts applied in 
the K$^+$K$^-$ analysis.}
\end{figure}

\noindent The meaning of this factor is:
if the matching efficiency for a particle is worse (better) in the experiment
than in the simulation, the factor is lower (higher) than 1. 
The $\theta_{\text{lab}} \text{--} p_{\text{T}}$ distribution of 
$\varepsilon^{CDC-PSB}$,
presented in Fig.~\ref{fig:bareff}, shows that discrepancies between the
simulated and experimental matching efficiencies remain 
mostly within a $\pm 8\%$ range. 
However, they arise at the detector edges and at low transverse momenta.
This finding was a motivation to limit the range of accepted
polar angles to $30^\circ < \theta_{\text{lab}} < 53^\circ$, 
and set the minimum $p_{\text{T}}$ at 0.18 (0.14) GeV/$c$ for K$^+$ (K$^-$), 
as mentioned above in Sect.~\ref{Sect:ana}.
Still, to correct for the remaining discrepancies, 
the kaon candidates selected to build the experimental
$M_{\text{inv}}$ spectrum were weighted with the factor 
\mbox{$1 \slash \varepsilon^{CDC-PSB} (\theta_{\text{lab}}, p_{\text{t}})$}.

More details on the estimation of systematic errors, conventions
applied and correction of the $\langle A_{\text{part}} \rangle_{\text{b}}$
value for the experiment presented in~\cite{Pias15} is reported in 
Appendix \ref{Sect:deta}.

\section{Results}
\label{Sect:resu}

\begin{table}
 \caption{Available data on the centrality dependence of the $\phi$ meson
 production yield per triggered event at 1.9A GeV. 
 Note, that some of published results were slightly adjusted
 according to the conventions defined in Appendix \ref{Sect:deta}.}
 \label{Tab:PhiYields}
 \begin{tabular}{cccc}
 \hline
 Reaction & $\langle A_{\text{part}} \rangle_{\text{b}}$ & Yield $\times 10^{-4}$ & Ref. \\
 \hline
  Al+Al   &   42.5 & $3.1 \pm 0.5 \text{(stat)} \pm 0.3 (\text{syst})$ & \cite{Gasi16} \\
  Ni+Ni   &   46.5 & $4.4 \pm 0.7 \text{(stat)} \pm 1.3 (\text{syst})$ & \cite{Pias15} \\
  Ni+Ni   &   74   & $8.6 \pm 1.6 \text{(stat)} \pm 1.5 (\text{syst})$ & This work \\
 \hline
  Ni+Ni (BUU) & 86 & 25. & \cite{Scha10} \\
 \hline
 \end{tabular}
\end{table}

\begin{figure}[b]
 \includegraphics[width=8.4cm]{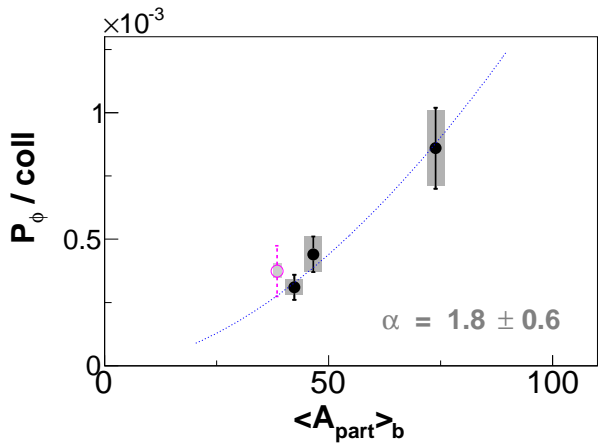}
 \caption{\label{fig:phiApart}(Color online)
  $\phi$ meson yield per event at the beam energy of 1.9A GeV
  as a function of number of participant nucleons averaged 
  over the impact parameters. 
  Full circles: data measured at 1.9A GeV; 
  open circle: data obtained at 1.756A GeV~\cite{Agak09}
  and scaled according to the systematics in~\cite{Meta93}; 
%full square: BUU simulation (cf.~\cite{Scha10}). 
  Dotted curve: power function fit 
  to the points marked by full circles. See text for details.
 }
\end{figure}

The production yield of $\phi$ mesons per triggered
collision of Ni+Ni ions was found to be:
\begin{equation}
 \rm{P(\phi) = [8.6 ~\pm~ 1.6 ~(\text{stat}) \pm 1.5 ~(\text{syst})] 
     \times 10^{-4}}\quad .
 \label{eq:yield_phi}
\end{equation}

\noindent where the systematic error is given within the 68.3\% 
confidence level (see Appendix~\ref{Sect:detaSystErr} for details).
For the consistency check we have verified that if the same 
centrality cut is imposed on the events of Ni+Ni collisions 
reported in~\cite{Pias15}, the $\phi$ meson yield extracted
from that data agrees within 1 standard deviation with the 
abovementioned result.

A compilation of the yields per event obtained by FOPI at 1.9A GeV 
is shown in Tab.~\ref{Tab:PhiYields} 
and in Fig.~\ref{fig:phiApart} (full circles). 
The multiplicity of $\phi$ mesons from Ar+KCl collisions at 1.756A GeV,
obtained by the HADES collaboration,
cannot be compared directly to this sample, 
as due to the scarcity of data on subthreshold $\phi$ meson production
the excitation function is not known. 
However, a common systematics of the excitation function of
$\pi^{0,-}$, K$^+$, $\eta$, and $\rho^0$ meson yields per participant
nucleon has been found, as shown in Fig.~6 of Ref.~\cite{Meta93}. 
Interestingly, if the $\phi$ meson yield from Ar+KCl collisions
is scaled using this systematics, the ''adjusted'' value 
shown as open circle in Fig.~\ref{fig:phiApart} falls well 
in line with the FOPI data points obtained at 1.9A GeV.

We have fitted the FOPI data with the power function,
\mbox{$P (\phi) \sim A_{\text{part}}^\alpha$}, 
and obtained $\alpha = 1.8 \pm 0.6$.
For the purpose of fitting, the statistical and systematic errors
were added quadratically. 
Within the thermal model approach without the strangeness undersaturation 
effects the volume dependence of the $\phi$ meson multiplicity in an event 
is predicted to be linear~\cite{Cley99}. 
The measured $\alpha$ parameter is in agreement with this prediction
within 1.4 standard deviations. 
According to the calculations in the frame of the BUU transport code
the $\phi$ meson yield in an event should be equal to $2.5 \times 10^{-3}$ 
for the class of 12\% most central Ni+Ni collisions
(corresponding to $\langle A_{\text{part}} \rangle_{\text{b}} = 86$ 
within the sharp cut-off approximation)~\cite{Scha10}. 
The experimental data for this centrality class (see~\cite{Mang03}) 
is too scarce for a precise comparison with this prediction. 
%However, the BUU result, shown as full square in Fig.~\ref{fig:phiApart},
However, the BUU result appears to overpredict the abovementioned 
power function, if the latter is slightly extrapolated beyond the region 
of experimental data points. 

The obtained centrality dependence of the $\phi$ meson production yield 
can be compared to the $A_{\text{part}}$ scaling of the other mesons 
produced at similar beam energies. 
The pattern for $\pi^{+,-}$ mesons exhibits somewhat 
sublinear dependence~\cite{Pelt97,Reis07}. 
For K$^0_{\text{s}}$, $\alpha = 1.20 \pm 0.25$ was reported~\cite{Mers07}.
Turning to the charged kaons, the centrality dependence of their
multiplicity was not measured at 1.9A GeV, 
but at 1.8A GeV the exponent was found to be
$\alpha = 1.65 \pm 0.15$ for K$^+$, and 
$\alpha = 1.8 \pm 0.3$ for K$^-$~\cite{Bart97}.

These observations can be interpreted by a mixture of different mechanisms. 
At beam energies around 1.9A GeV charged pions are produced 
abundantly far above the threshold in the free NN collision, 
with considerable feeding from $\Delta$ resonance decays~\cite{Hong97}. 
Their sublinear centrality dependence was interpreted 
as the result of interplay between the surface ($\sim \text{A}^{2/3}$) 
and volume ($\sim$A) contributions~\cite{Reis07}. 
On the other hand the analysis of $\pi^+$ mesons emitted from 
the collisions around \mbox{0.8--1.0A~GeV}
has shown that whereas for pions with energies allowed in the
free NN channel the $\alpha$ exponent is consistent \mbox{with 1}, 
for high energy pions it exceeds the unity. In the latter domain
the exponent rises with the pion energy, 
as more and more additional energy is necessary to pass the
threshold in the NN channel (see Fig.~11 in Ref.~\cite{Muen97}). 

The rise of $\alpha$ coefficient with probing more subthreshold energies 
was also observed for the K$^+$ production~\cite{Bart97,Fors07}. 
At the beam energies of 1--2A GeV these particles are produced
around the threshold in the $\text{NN} \rightarrow \text{NKY}$ channel,
which is energetically the lowest-lying channel 
for the K$^+$ production in the free NN collision. 
Therefore the multi-step channels involving the intermediate 
$\Delta$ baryons or pions, which usually facilitate the subthreshold
production of particles, are the important competitors, 
and -- according to the BUU and IQMD transport codes -- 
these channels are responsible for 
over a half of the kaon production cross section~\cite{Scha10,Hart11}. 
Interestingly, within the thermal model approach the strong rising trend 
of K$^+$ yield in an event with number of participants 
is explained by the associated production of a hyperon in the
$\text{NN} \rightarrow \text{NKY}$ channel, required
to conserve the strangeness~\cite{Cley99}. 
Regarding the K$^-$ production, at beam energies between 1 and 2A GeV 
they are produced quite deeply below the threshold in the free NN
collision, and the centrality dependence of K$^-$ multiplicity in an event 
was found to be stronger than linear, repeating the case of K$^+$ and pions. 
In addition, for a fixed energy the centrality dependence of K$^-$ yield 
appeared to follow the pattern for K$^+$~\cite{Bart97,Fors07}. 
The latter observation can be explained by noticing that the 
strangeness-exchange channels ($\pi\text{Y} \rightarrow \text{BK}^-$ and
$\text{BY} \rightarrow \text{BBK}^-$), 
which are expected to dominate the K$^-$ production
(cf.~\cite{Fuch06,Scha10,Hart11}), require the presence of a hyperon, while 
the energetically lowest-lying channel leading to the hyperon production,
$\text{NN} \rightarrow \text{NYK}^{+,0}$, generates also the kaon.

Turning back to the $\phi$ meson production, the extracted value 
of the exponent is characterized by the large uncertainty. 
Therefore, at the present level of statistics, the production mechanism
cannot be judged from the comparison of $\alpha$ parameters 
between $\phi$ and other mesons. 
Interestingly, the state-of-the-art calculations in the
frame of the BUU and UrQMD models advocate the entirely different
mechanisms of the $\phi$ meson production~\cite{Scha10,Stei16}. 

\begin{figure}[b]
 \includegraphics[width=8.6cm]{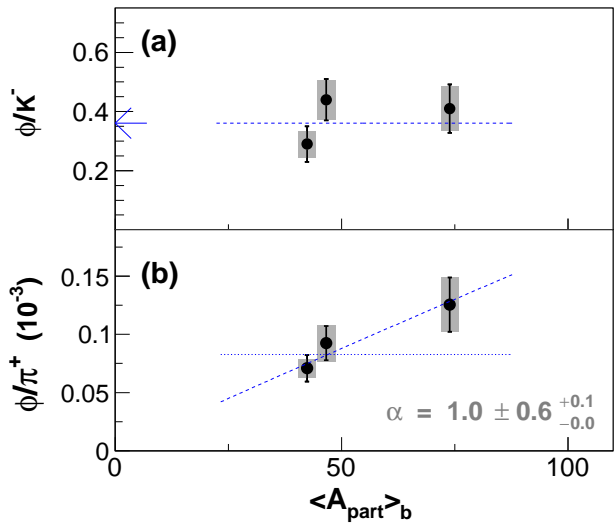}
 \caption{\label{fig:phikmpip}(Color online) 
  (a) Centrality dependence of \mbox{$\phi\slash\text{K}^-$} ratio
      at the beam energy of 1.9A GeV.
      The horizontal line and the arrow mark the average value.
  (b) Centrality dependence of \mbox{$\phi\slash\pi^+$} ratio. 
      The dotted line shows the fit of the constant function, and
      the dashed line -- the power function fit. 
      The value of $\alpha$ exponent is the result of the latter fit. 
      See text for details.
}
\end{figure}

Considering the ratio of \mbox{$\phi\slash\text{K}^-$} yields, 
in case of the central Ni+Ni collisions, 
we notice that the sample of negative kaons reported in~\cite{Menz00}
was measured within nearly the same centrality class as 
$\phi$ mesons from the S261 experiment. 
To obtain the production yield per event of these kaons, 
we have integrated the rapidity distribution shown in Fig.~2 of 
Ref.~\cite{Menz00}. It allowed us to find the ratio of yields: 

\begin{equation}
 \rm{\frac{P(\phi)}{P(\text{K}^-)} = 0.41 \pm 0.08 (stat) \pm 0.08 (syst) 
     \quad ,} 
\end{equation}

\noindent where the systematic error is given within the 68.3\%
confidence level and includes the relevant uncertainties associated 
to both the experiments. With this information 
we have plotted the centrality dependence of the 
\mbox{$\phi\slash\text{K}^-$} ratio, as shown in Fig.~\ref{fig:phikmpip}(a).
No clear trend is observed. The average value of this ratio is 
$0.36 \pm 0.05$. Taking into account the branching ratio of 
the $\phi \rightarrow \text{K}^+\text{K}^-$ decay we conclude
that at the beam energy of 1.9A GeV,
$(18 \pm 3)$ \% of negative kaons originate from the decays 
of $\phi$ mesons, in agreement with the previous analyses~\cite{Pias15,Gasi16}.
Interestingly, these results are the same as for the Ar+KCl case, 
measured at the more subthreshold beam energy of 1.756A GeV~\cite{Agak09}. 
Within the version of statistical model in which the canonical ensemble 
is applied to particles with open strangeness, the studied ratio is 
sensitive to the correlation length $R_{\text{C}}$, 
which corresponds to the reduced volume where strangeness can be produced.
As shown in Fig.~13 of Ref.~\cite{Agak09}, this sensitivity becomes
particularly strong around the $\phi$ and K$^-$ threshold energies.
The value of \mbox{$\phi\slash\text{K}^-$}
ratio obtained in the present analysis, as well as from the abovementioned
Ar+KCl collisions fall within the range of $R_{\text{C}} \in (2.2, 3.2)$~fm. 
Most recent calculations within the UrQMD model appear to reproduce the found 
\mbox{$\phi\slash\text{K}^-$} ratio as well (cf. fig.~4a in~\cite{Stei16}).

Fig~\ref{fig:phikmpip}(b) shows the centrality dependence of the 
\mbox{$\phi\slash\pi^+$} ratio (see Appendix~\ref{Sect:detaCenKmPi} 
for more details). 
We have fitted the data with the constant function 
(shown as dotted line) and obtained the value 
$(8.3 \pm 1.1) \times 10^{-5}$ at \mbox{$\chi^2\slash\text{NDF} = 1.3$}. 
An alternative fit of the power function,
\mbox{$P (\phi) \slash P(\pi^+) \sim A_{\text{part}}^\alpha$}, 
yields $\alpha = 1.0 \pm 0.6 ^{+0.1}_{-0.0}$ at 
\mbox{$\chi^2\slash\text{NDF}$}~=~0.3. 
A minor systematic error accounts for some uncertainty of the 
$\pi^+$ yield at $\langle A_{\text{part}} \rangle_{\text{b}} = 42.5$, 
which is discussed in Appendix~\ref{Sect:detaCenKmPi}.

The statistical model predicts the \mbox{$\phi\slash\pi^+$} ratio to be
independent of the number of participant nucleons
(see Fig.~2 in~\cite{Cley99}). 
On the other hand, according to the most recent calculations within 
the UrQMD model, the predicted trend of the abovementioned ratio,
shown in fig.~4b in Ref.~\cite{Stei16}, is strongly rising with $A_{\text{part}}$.
Judging by the obtained values of $\chi^2\slash\text{NDF}$, both
scenarios are consistent with our data.
Within the latter model the \mbox{$\phi\slash\pi^+$} 
ratio is predicted to be sensitive to the temperature of the collision zone, 
which by comparison to the abovementioned figure in~\cite{Cley99}, 
would be estimated to be around 70~MeV. 

\section{Summary}
We have investigated the $\phi$ meson production in the K$^+$K$^-$ 
decay channel in the 22\% most central collisions of Ni+Ni 
at the beam kinetic energy of 1.93A GeV 
and found the sample of $110 \pm 19$ mesons. 
After the efficiency corrections the yield per triggered event was found to be 
$[8.6 ~\pm~ 1.6 ~(\text{stat}) \pm 1.5 ~(\text{syst})] \times 10^{-4}$.
It allowed for the first time to gain insight into the centrality 
dependence of the multiplicity of $\phi$ mesons produced 
in the heavy-ion collisions below the NN threshold. 
Its dependence can be described by a power law with the exponent
$\alpha$ of \mbox{$1.8 \pm 0.6$}.
The current BUU calculations seem to overpredict this trend.
The \mbox{$\phi\slash\text{K}^-$} ratio as a function of $A_{\text{part}}$
does not exhibit any trend within the available accuracy, 
and the average value was found to be $0.36 \pm 0.05$. 
We have also found that the \mbox{$\phi\slash\pi^+$} ratio 
has an average value of $(8.3 \pm 1.1) \times 10^{-5}$. 
A power function parameterization of the $A_{\text{part}}$ dependence 
of this ratio yielded the exponent of $\alpha = 1.0 \pm 0.6 ^{+0.1}_{-0.0}$.
According to the values of $\chi^2\slash\text{NDF}$ for these fits,
both of them are consistent with the data.

\begin{table}
 \caption{Variations of the cut parameters used in the investigation of
          $\phi$ mesons in the central Ni (1.9A GeV) + Ni collisions}
 \label{Tab:Syst}
 \begin{ruledtabular}
  \begin{tabular}{p{1.7cm}|p{4.6cm}|p{3cm}}

Parameter 
          & Meaning 
          & Values \\
\hline
$T$ [MeV] & Temperature of $\phi$ meson source
          & 80,~100, ...,~140 \\
$a_2$     & Anisotropy parameter of $\phi$ meson source 
          & {0.0 , 1.0} \\
$M_{\text{inv,norm}}^{\text{min}}$ [GeV/$c^2$] 
          & Lower endpoint of range of normalisation of background to true pairs 
          & 1.05, 1.10 \\
$M_{\text{inv,int}}^{[\text{min--max}]}$ [GeV/$c^2$] 
          & Integration window of $\phi$ meson signal in the $M_{\text{inv}}$ spectrum 
          & \mbox{[0.98~--~1.05]}, \mbox{[1.00~--~1.04]}, \mbox{[1.01~--~1.04]} \\
$N_{\text{bins}}$ 
          & No. of bins for $M_{\text{inv}}$ spectrum 
          & {25, 35, ..., 65} \\
$hmul_{[\text{K}^+, \text{K}^-\!]}^{\text{min}}$
          & Minimum multiplicity of hits in the CDC for a kaon candidate
          & \mbox{[27, 27]}, \newline \mbox{[30, 31]} \\
$p_{\text{K}^-}^{\text{max}}$ [GeV/$c$]
          & Maximum momentum of K$^-$ candidate & 0.60, 0.65 \\

  \end{tabular}
 \end{ruledtabular}
\end{table}

\appendix
\section{More details on the analysis}
\label{Sect:deta}

\subsection{Systematic errors}
\label{Sect:detaSystErr}

Table~\ref{Tab:Syst} shows the different ranges of parameters used 
in the present $\phi$ meson analysis. 
Since none of the parameter combinations resulted in 
a significantly different value of the $\phi$ meson yield, 
the latter has been determined by averaging the contributions
from these combinations, as detailed in Eq.~\ref{eq:yield_phi}. 
This approach also allowed to choose the confidence level (CL)
at which the systematic errors were to be extracted. 
In this work we took the range containing 68.3\% 
of the abovementioned contributions, 
summed up around the mean value of the yield. 

\subsection{Centrality dependence of $\phi$ yield per event}
\label{Sect:detaCen}

For the analysis of the centrality dependence of the multiplicity
of $\phi$ mesons at 1.9A GeV, 
we include the data from semi-central Ni+Ni collisions presented 
in~\cite{Pias15} and from central collisions of Al+Al~\cite{Gasi16}.
We have reanalysed the centrality determination procedure for the
data published in~\cite{Pias15}, and got a slightly corrected value
of the cross section obtained with the used trigger: 
$(56 \pm 3)$\% of the total reaction cross section. 
Within the geometrical model of interpenetrating spheres 
this translates into the average number of participants of 
$\langle A_{\text{part}} \rangle_{\text{b}} = 46.5 \pm 2.0$.

In order to stay consistent with the method used to determine
the yield value and systematic errors described in 
Sect.~\ref{Sect:detaSystErr},
the data from~\cite{Pias15} and~\cite{Gasi16} had to be adjusted. 
For the Ni+Ni analysis reported in~\cite{Pias15}, 
the "CL = 68.3\%" systematic error was obtained by reanalysing 
the distribution of possible values of the $\phi$ meson multiplicity
per event resulting from various combinations of the applied cuts. 
In case of the Al+Al analysis presented in \cite{Gasi16},
we have shifted the value of yield slightly 
to remove the asymmetry of systematic errors. 
In addition, as the systematic errors for the latter analysis
corresponded to \mbox{$\text{CL} > 95\%$} range, 
we took the published value of this error divided by two 
as an upper estimation of the \mbox{$\text{CL} = 68.3\%$} value.

\subsection{Centrality dependence of $\phi\slash\text{K}^-$ and 
\mbox{$\phi\slash\pi^+$} ratios}
\label{Sect:detaCenKmPi}

To study the centrality dependence of the \mbox{$\phi\slash\text{K}^-$} 
ratio we used the published values from~\cite{Pias15} and~\cite{Gasi16}
and applied slight corrections according to the prescriptions 
specified above in the Appendix~\ref{Sect:detaCen}. 
To reconstruct the centrality dependence of the \mbox{$\phi\slash\pi^+$} 
ratio, for two samples of Ni+Ni data
we interpolated the data points from the $A_{\text{part}}$ dependence
of the $\pi^+$ yield per event, reported in Fig.~1 of Ref.~\cite{Pelt97}. 
In case of the Al+Al collisions, the $\pi^+$ yield is not yet known. 
As an estimation of this yield we took the yield of positive pions
from Ni+Ni at the same number of participant nucleons as for 
the Al+Al experiment. We motivate this substitution by three arguments:
\begin{itemize}

  \item at beam energies of 1 and 1.5A GeV the pion yield 
  {\it per number of participant nucleons} as function of 
  $A_{\text{part}}$ was found to be parameterized by one (universal) 
  curve for a wide range of system sizes, as shown in the upper panels of 
  Fig.~18 of Ref.~\cite{Reis07}.

  \item the N/Z ratios for Al+Al and Ni+Ni systems are the same,

  \item the pion yield per $A_{\text{part}}$ 
  for the collisions of the moderate-size Ni+Ni system at 1A GeV was found 
  to be 15\% higher than that for the collisions of large Au+Au 
  system (see p.~480 in Ref.~\cite{Reis07}). If we assume that
  the further drop of the system size from Ni+Ni to Al+Al generates 
  the same rise of the pion yield per $A_{\text{part}}$, 
  it introduces a change of the value of the fitted $\alpha$
  exponent from 1.0 to 1.1, which is minor compared to the fit error 
  associated to this parameter. Therefore we introduce an additional
  systematic error of +0.1 due to this uncertainty. 
\end{itemize}

\begin{acknowledgments}
This work was supported in part by the German Federal Ministry of Education 
and Research (BMBF) under grant No. 06HD154, 
by the Gesellschaft f\"ur Schwerionenforschung (GSI) under grant No. HD-HER,
by the Korea Research Foundation grant (KRF-2005-041-C00120), 
by the European Commision under the 6th Framework Programme under the
'Integrated Infrastructure Initiative on: Strongly Interacting Matter 
(Hadron Physics)' (Contract No. RII3-CT-2004-506078)
and by the agreement between GSI and IN2P3/CEA.
\end{acknowledgments}

\end{document}